\documentclass[epj,final,amsmath,amssymb]{svjour}
\usepackage{graphicx}
\usepackage{amssymb}

\input{epsf}

\begin{document}

\title{Google matrix of Twitter}

\author{Klaus M. Frahm and Dima L. Shepelyansky}

\institute{Laboratoire de Physique Th\'eorique du CNRS, IRSAMC, 
Universit\'e de Toulouse, UPS, F-31062 Toulouse, France
\and
http://www.quantware.ups-tlse.fr}

\titlerunning{Google matrix of Twitter}
\authorrunning{K.M.Frahm and D.L.Shepelyansky}

\date{July 14, 2012}
%\date{\today}

\abstract{We construct the Google matrix of the entire Twitter network,
dated by July 2009, and analyze its spectrum and eigenstate properties
including the PageRank and CheiRank vectors and 2DRanking of all nodes.
Our studies show much stronger inter-connectivity between top PageRank nodes
for the Twitter network compared to the networks of Wikipedia
and British Universities studied previously. Our analysis allows 
to locate the top Twitter users which control the information
flow on the  network. We argue that this small
fraction of the whole number of users, which can be viewed as  the social network elite,
plays the dominant role in the process of opinion formation on the network.
}

\maketitle
%%%%%%%%%%%%%%%%%%%%%%%%%%%%%%%%%%%%%%%%%%%%%%%%%%%%%%%%%
\section{Introduction}
\label{sec1}

Twitter is an online directed social network 
that enables its users to exchange short communications
of up to 140 characters \cite{wikitwitter}.
In March 2012 this network had around 140 million active users
\cite{wikitwitter}. Being founded in 2006, the size of this network
demonstrates an enormously fast growth 
with 41 million users in July 2009 \cite{twitter2009}, 
only three years after its creation. 
The crawling and statistical analysis
of the entire Twitter network, collected in July 2009, was done 
by the KAIST group \cite{twitter2009} with additional statistical
characteristics available at LAW DSI of Milano University 
\cite{twittervigna}. This network has scale-free properties
with an average power law distribution  of ingoing and outgoing links
\cite{twitter2009,twittervigna} being typical for the World Wide Web (WWW), 
Wikipedia and other social networks (see e.g \cite{donato}, 
\cite{upfal}, \cite{wiki}). In this work we use this Twitter dataset to 
construct the Google matrix \cite{brin,meyerbook}
of this directed network and we analyze the spectral properties of 
its eigenvalues and eigenvectors. Even if the entire size of Twitter 2009
is very large the powerful Arnoldi method (see e.g. \cite{arnoldibook}, 
\cite{golub}, \cite{ulamfrahm}, \cite{univpr})
allows to obtain the spectrum and eigenstates for the largest eigenvalues.

A special analysis is performed for the PageRank vector, used in the 
Google search engine \cite{brin,meyerbook}, and the Chei\-Rank vector 
studied for the Linux Kernel network \cite{alik,wlinux}, 
Wiki\-pedia articles network \cite{wiki},
world trade network \cite{wtrade} and other directed networks \cite{2dmotor}.  
While the components of the  PageRank vector 
are on average proportional to a number of ingoing links \cite{litvak},
the components of the CheiRank vector are 
on average proportional to a number of outgoing links \cite{wiki,alik}
that leads to a two-dimensional ranking of all network nodes \cite{2dmotor}.
Thus our studies allow to analyze the spectral properties of the entire 
Twitter network of an 
enormously large size which is by one-two orders of magnitude larger
compared to previous studies \cite{wiki,univpr,wlinux,2dmotor}.

The paper is organized as follows: the construction of the Google matrix and 
its global structure are described in Section 2; the properties of 
spectrum and eigenvectors of the Google matrix of Twitter are 
presented in Section 3;
properties of 2DRanking of Twitter network are analyzed in Section 4
and the discussion of the results is given in Section 5.
Detailed data and results of our statistical analysis 
of the Twitter matrix are presented at the web page 
\cite{twittermatrix}.

%%%%%%%%%%%%%%%%%%%%%%%%%%%%%%%%%%%%%%%%%%%%%%%%%%%%%%%%%
\section{Google matrix construction}
\label{sec2}

The Google matrix of the Twitter network is constructed following the 
standard rules described in \cite{brin,meyerbook}: we consider the elements  
$A_{ij}$ of the adjacency matrix
being equal to unity if a user (or node) 
$j$ points to user $i$ and zero otherwise. Then the Google matrix 
of the network with  $N$ users is given by
\begin{equation}
  G_{ij} = \alpha S_{ij} + (1-\alpha) / N \;\; ,
\label{eq1} 
\end{equation} 
where the matrix $S$ is obtained by normalizing 
to unity all columns of the adjacency matrix $A_{i,j}$ with at least 
one non-zero element, and replacing columns with only zero elements, 
corresponding to the dangling nodes, by $1/N$.
The damping factor $\alpha$ in the WWW context describes the probability 
$(1-\alpha)$ to jump to any node for a random surfer. 
The value $\alpha = 0.85$  gives
a good classification for WWW \cite{meyerbook}
and  thus we also use this value here.
The matrix $G$ belongs to the class of Perron-Frobenius 
operators \cite{meyerbook},
its largest eigenvalue 
is $\lambda = 1$ and other eigenvalues have $|\lambda| \le \alpha$. 
The right eigenvector 
at $\lambda = 1$ gives the probability $P(i)$ to find 
a random surfer at site $i$ and
is called the PageRank. Once the PageRank is found, 
all nodes can be sorted by decreasing probabilities $P(i)$. 
The node rank is then given by index $K(i)$ which
reflects the  relevance of the node $i$. The top 
PageRank nodes are located at small values of $K(i)=1,2,...$.

\begin{figure}[h!]
\begin{center}
\includegraphics[width=0.48\textwidth]{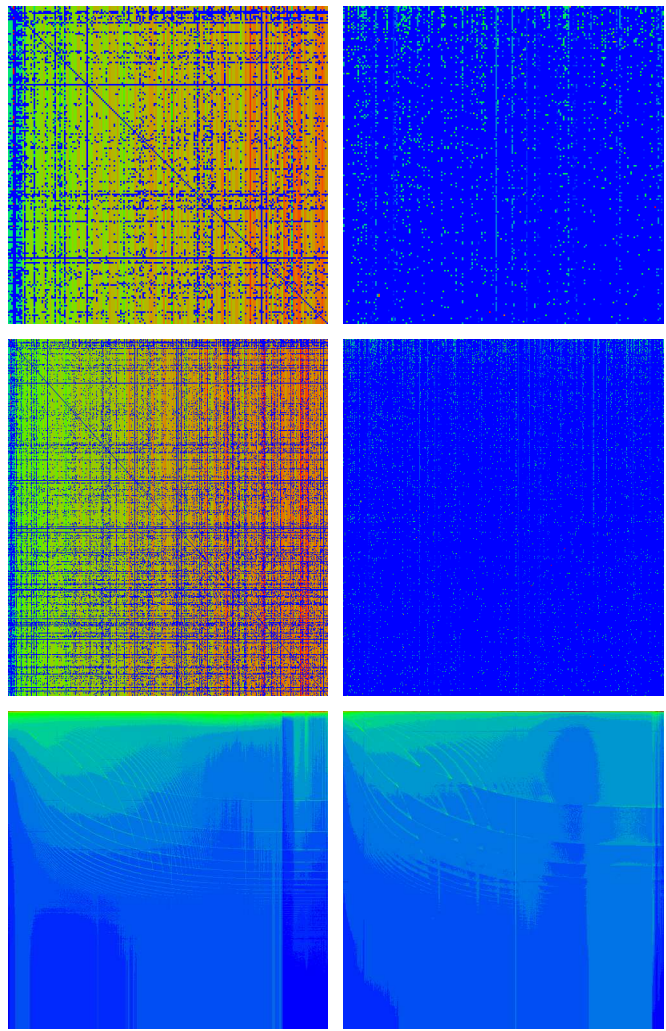}
\end{center}
\caption{\label{fig1} 
Google matrix of Twitter: matrix elements of $G$ (left column)
and $G^*$ (right column) are shown in the basis of
PageRank index $K$ (and $K'$) of matrix $G_{KK'}$ (left column panels)
and in the basis of CheiRank index $K^*$ (and $K^{*'}$)
of matrix ${G^*}_{{K^*}{K^{*'}}}$ (right column panels).
Here, $x$ (and $y$) 
axis show $K$ (and $K^{'}$) (left column)
(and respectively $K^{*}$ and $K^{*'}$ on right column)
with the range 
$1 \leq K,K' \leq 200$ (top panels);
$1 \leq K,K' \leq 400$ (middle panels);
$1 \leq K,K' \leq N$ (bottom panels).
All nodes are ordered by PageRank index $K$ of the matrix $G$ and thus we 
have two matrix indexes
$K,K'$ for matrix elements in this basis (left column) and
respectively $K^*,K^{*'}$ for matrix $G^*$ (right column).
Bottom panels show the 
coarse-grained density of matrix elements $G_{K,K'}$ and 
$G^*_{{K^*}{K^{*'}}}$; 
the coarse graining is done on $500 \times 500$ square cells 
for the entire Twitter network.
We use a standard matrix representation with $K=K'=1$ on 
top left panel corner (left column) and
respectively $K^{*}=K^{*'}=1$ (right column).
Color shows the amplitude of matrix elements
in top and middle panels or their density in the bottom panels
changing from blue for minimum zero value to red at maximum value.
Here the damping factor is $\alpha=1$.
}
\end{figure}

The PageRank dependence on $K$ is well
described by a power law $P(K) \propto 1/K^{\beta_{in}}$ with
$\beta_{in} \approx 0.9$. This is consistent with the relation
$\beta_{in}=1/(\mu_{in}-1)$ corresponding to the average
proportionality of PageRank probability $P(i)$
to its in-degree distribution $w_{in}(k) \propto 1/k^{\mu_{in}}$
where $k(i)$ is a number of ingoing links for a node $i$  
\cite{meyerbook,litvak}. For the WWW it is established that
for the ingoing links $\mu_{in} \approx 2.1$ (with $\beta_{in} \approx 0.9$)
while for the out-degree distribution
$w_{out}$ of
outgoing links the power law has the exponent  $\mu_{out} \approx 2.7$
\cite{donato,upfal}.
Similar values of these exponents are found for 
the WWW British university networks \cite{univpr}, 
the procedure call network 
of Linux Kernel software introduced in \cite{alik}
and for Wikipedia hyperlink citation network of English articles 
%with corresponding $\beta_{out}=1/(\mu_{out}-1) \approx 0.6$
(see e.g. \cite{wiki}). 

In addition to the Google matrix $G$ we also analyze the properties of 
matrix $G^*$ constructed from the network with inverted
directions of links, with the adjacency matrix $A_{i,j} \rightarrow A_{j,i}$.
After the inversion of links the Google matrix $G^*$ is constructed 
via the procedure (\ref{eq1}) described above.
The right eigenvector at unit eigenvalue of the matrix $G^*$ is called
the CheiRank \cite{alik,wiki}. 
In analogy with the PageRank the probability values of CheiRank
are proportional to number of outgoing links, due to links inversion.
All nodes of the network can be ordered in a decreasing order
with the CheiRank index $K^*(i)$ with $P^* \propto 1/{K^*}^{\beta_{out}}$
with $\beta_{out}=1/(\mu_{out}-1)$. Since each node $i$ of the network 
is characterized both by PageRank $K(i)$ and CheiRank $K^*(i)$ indexes
the ranking of nodes becomes two-dimensional. While PageRank
highlights well-know popular nodes, CheiRank highlights
communicative nodes. As discussed in \cite{wiki,alik,2dmotor},
such 2DRanking allows to characterized 
an information flow on networks in a more efficient and rich manner.
It is convenient to characterize the interdependence between 
PageRank and CheiRank vectors by the correlator 
\begin{equation}
  \kappa =N \sum^N_{i=1} P(K(i)) P^*(K^*(i)) - 1 \;\; .
\label{eq2} 
\end{equation}
As it is shown in \cite{alik,2dmotor}, 
we have $\kappa \approx 0$ 
for Linux Kernel network, transcription gene networks
and $\kappa \approx 2 - 4$ for University and Wikipedia networks.

In this work we apply the Google matrix analysis
developed in \cite{wiki,univpr,alik,wlinux,wtrade,2dmotor}
to the Twitter 2009 network available at \cite{twitter2009,twittervigna}.
The total size of the Google matrix is $N=41652230$
and the number of links is $N_\ell=1468365182$.
This matrix size is by one-two orders of magnitude larger
than those studied in \cite{univpr,wlinux,2dmotor}.
The number of links per node is $\xi_\ell =N_\ell/N \approx 35$
being by a factor $1.5 - 3.5$ larger than for Wikipedia network or
Cambridge University 2006 network \cite{2dmotor}.
The matrix elements of $G$ and $G^*$ are shown in Fig.~\ref{fig1}
on a scale of top 200 (top panels) and 
400 (middle panels) values of $K$ (for $G$) and $K^*$ (for $G^*$)
and in a coarse grained image for the whole matrix size scale (bottom panels).

It is interesting to note that the coarse-grained image has well visible hyperbolic onion 
curves of high density which are similar to those
found in  \cite{2dmotor} for Wikipedia and University networks.
In \cite{2dmotor} the appearance of such curves was attributed to
existence of specific categories. We assume that for the Twitter
network such curves are a result of enhanced links between
various categories of users (e.g. actors, journalists etc.)
but a detailed origin is still to be established. 

In the following sections we also compare the properties of the Twitter 
network with those of the Wikipedia articles network
from \cite{wiki}. Some spectral properties of the Wikipedia network 
with $N=3282257$ nodes and $N_\ell=71012307$ links are analyzed in 
\cite{univpr,2dmotor}.
We also compare certain parameters with the 
networks of Cambridge and Oxford Universities of 2006
with $N=212710$ and $N=200823$ nodes and 
with $N_\ell=2015265$ and $N_\ell=1831542$ links respectively.
The properties of these networks are discussed in
\cite{univpr,2dmotor}. The gallery of the Google matrix $G$ images for these networks,
as well as for the Linux Kernel network, are presented in \cite{2dmotor}.
The comparison with the data shown in Fig.~\ref{fig1} here
shows that for the Twitter network we have much stronger interconnection matrix
at moderate $K$ values. We return to this point in Sections 4,5.

%%%%%%%%%%%%%%%%%%%%%%%%%%%%%%%%%%%%%%%%%%%%%%%%%%%%%%%%%
\section{Spectrum and eigenstates of Twitter}
\label{sec3}

To obtain the spectrum of the Google matrix of Twitter
we use the Arnoldi method \cite{arnoldibook,golub,ulamfrahm}.
However, at first, following the approach developed in \cite{univpr},
we determine the invariant subspaces of the Twitter network.
For that for each node we find iteratively the set of nodes 
that can be reached by a chain of non-zero matrix elements of $S$. 
Usually, there are several such invariant isolated subsets and
the size of such subsets is smaller than the whole matrix size.
These subsets are invariant with respect to applications of matrix $S$.
We merge all subspaces with common 
members, and obtain a sequence of disjoint subspaces $V_j$ of dimension 
$d_j$ invariant by applications of $S$. The remaining part of nodes 
forms the wholly connected {\it core space}. 
Such a classification scheme can be efficiently implemented in a 
computer program, it provides a subdivision of network nodes  in $N_c$ 
core space nodes (typically 70-80\% of $N$ for British University networks
\cite{univpr}) and $N_s$ subspace nodes 
belonging to at least one of the invariant subspaces $V_j$ 
inducing the block triangular structure, 
\begin{equation}
\label{eq3}
S=\left(\begin{array}{cc}
S_{ss} & S_{sc}  \\
0 & S_{cc}\\
\end{array}\right)\; .
\end{equation}
Here the subspace-subspace block $S_{ss}$ is actually composed of 
many diagonal blocks for each of the invariant subspaces. Each of these blocks 
corresponds to a column sum normalized matrix of the same type as $G$ 
and has therefore at least one unit eigenvalue thus explaining the 
high degeneracy. Its eigenvalues and eigenvectors are easily accessible by 
numerical diagonalization (for full matrices) thus allowing to count the 
number of unit eigenvalues. 

\begin{figure}[h]
\begin{center}
\includegraphics[width=0.48\textwidth]{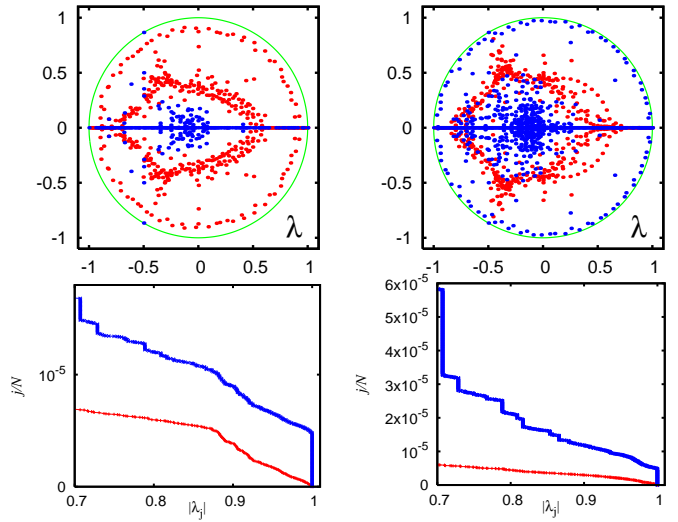}
\end{center}
\caption{\label{fig2} 
Spectrum of the Twitter matrix $S$ ($S^*$ with inverted direction of links) 
for the Twitter network shown on left panels (right panels). 
{\em Top panel:} Subspace eigenvalues (blue dots)  and 
core space eigenvalues (red dots) in $\lambda$-plane 
(green curve shows unit circle);
there are 17504 (66316) invariant subspaces, with maximal dimension 44 
(2959) and the sum of all subspace dimensions is $N_s=40307$ (180414).
The core space eigenvalues are obtained from the Arnoldi method applied to 
the core space subblock $S_{cc}$ of $S$ with Arnoldi dimension 640 as 
explained in Ref.~\cite{univpr}. 
{\em Bottom panels:} 
Fraction $j/N$ of eigenvalues with $|\lambda| >  |\lambda_j|$ for the 
core space eigenvalues (red bottom curve) and all eigenvalues (blue top curve)
from raw data of top panels. The number of 
eigenvalues with $|\lambda_j|=1$ is 34135 (129185) of which 
17505 (66357) are at $\lambda_j=1$; this number is (slightly) larger than the 
number of invariant subspaces which have each at least one unit eigenvalue.
Note that in the bottom panels the number of eigenvalues with $|\lambda_j|=1$ is 
artificially reduced to 200 in order to have a better scale on the 
vertical axis. The correct number of those eigenvalues corresponds to 
$j/N=8.195\times 10^{-4}$ ($3.102\times 10^{-3}$) which is strongly outside 
the vertical panel scale. 
}
\end{figure}

We find for the $G$ matrix of Twitter 2009
that there are $N_s=40307$ subset sites with a maximal subspace dimension 
of 44 (most subspaces are of dimension 2 or 3). For the matrix $G^*$ we find
$N_s=180414$ also with a lot of subspaces of dimension 2 or 3 and 
a maximal subspace dimension of 2959. 
The remaining eigenvalues of $S$ can be obtained from 
the projected core block 
$S_{cc}$ which is not column sum normalized 
(due to non-zero matrix elements 
in the block $S_{sc}$) and has therefore eigenvalues 
strictly inside the 
unit circle $|\lambda^{\rm (core)}_j|<1$. We have applied the 
Arnoldi method (AM) 
\cite{arnoldibook,golub,ulamfrahm} with Arnoldi dimension $n_A=640$ 
to determine the largest eigenvalues of $S_{cc}$ which required 
a machine with 250 GB of physical RAM memory to store the non-zero 
matrix elements of $S$ and the 640 vectors of the Krylov space.

In general the Arnoldi methods provides numerically accurate values 
for the largest eigenvalues (in modulus) but their number 
depends crucially on the Arnoldi dimension. In our case there is a 
considerable density of real eigenvalues close to the points $1$ and 
$-1$ where convergence is rather difficult. Comparing the results for 
different values of $n_A$, we find that for the matrix $S$ ($S^*$) the 
first 200 (150) eigenvalues are correct within a relative error below 
$0.3\;\%$ while the marjority of the remaining eigenvalues with 
$|\lambda_j|\ge 0.5$ ($|\lambda_j|\ge 0.6$) have a relative error of 
$10\;\%$. However, the well isolated complex eigenvalues, well visible in 
Fig.~\ref{fig2}, converge much better and are numerically accurate 
(with an error $\sim 10^{-14}$). The first three core space eigenvalues 
of $S$ ($S^*$) are also numerically acurrate with an error of $\sim 10^{-14}$ 
($\sim 10^{-8}$). 

The composed
spectrum of subspaces and core space eigenvalues obtained by the Arnoldi method
is shown in Fig.~\ref{fig2} for $G$ and $G^*$.  
The obtained results show that the fraction of invariant subspaces 
with $\lambda=1$ ($g_1= N_s/N \approx 10^{-3}$) is by orders of magnitude 
smaller than the one found for British Universities ($g_1 \approx 0.2$ at 
$N \approx 2 \times 10^5$) \cite{univpr}. 
We note that the cross and triple-star structures are visible for
Twitter spectrum in Fig.~\ref{fig2} but they
are significantly less pronounced as compared to the case of Cambridge and 
Oxford network spectrum (see Fig.2 in \cite{univpr}). It is interesting 
that such a triplet and cross structures naturally appear in the 
spectra of random unistochastic matrices of size $N=3$ and $4$ which 
have been analyzed analytically and numerically in \cite{karol2003}.
A similar star-structure spectrum appears also in sparse regular graphs 
with loops
studied recently in \cite{bolle} even if in the later case 
the spectrum goes outside of unit circle.
This shows that even in large size networks the loop structure between 
$3$ or $4$ dominant types of nodes is well visible for University networks.
For Twitter network it is less pronounced probably due to a larger number
$\xi_\ell$ of links per node. At the same time a circle structure in the spectrum remains
well visible both for Twitter and University networks. 
The integrated number of eigenvalues as a function of $|\lambda|$
is shown in the bottom panels of Fig.~\ref{fig2}.
Further detailed analysis is 
required for a better understanding of the origin of such spectral structures.

\begin{figure}[h]
\begin{center}
\includegraphics[width=0.48\textwidth]{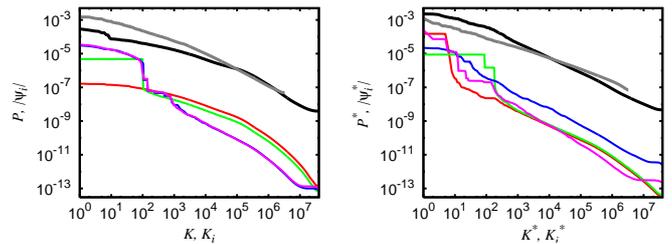}
\end{center}
\caption{\label{fig3} 
The left (right) panel shows the PageRank $P$ (CheiRank $P^*$)
versus the corresponding rank index $K$ ($K^*$)
for the Google matrix of Twitter at the damping parameter 
$\alpha=0.85$ (thick black curve); for comparison the PageRank (CheiRank)
of the Google matrix of Wikipedia network 
\cite{wiki} is shown by the gray curve at same $\alpha$.  
The colored thin curves (shifted down by factor 1000 for clarity) show 
the modulus of four core space eigenvectors $|\psi_i|$ ($|\psi_i^*|$)
of $S$ ($S^*$) versus their own ranking 
indexes $K_i$ ($K_i^*$).
Red and green lines are the eigenvectors corresponding to 
the two largest core space eigenvalues (in modulus) 
$\lambda_1=0.99997358$, $\lambda_2=0.99932634$ 
($\lambda_1=0.99997002$, $\lambda_2=0.99994658$); blue and pink lines 
are the eigenvectors corresponding to the two complex eigenvalues 
$\lambda_{151}=0.09032572+i\,0.90000530$, 
$\lambda_{161}=-0.47504961+i\,0.76576321$
($\lambda_{457}=0.38070896+i\,0.39207668$, 
$\lambda_{105}=-0.45794117+i\,0.80825210$). Eigenvalues and eigenvectors 
are obtained by the Arnoldi method with Arnoldi dimension 640 as for the data 
in Fig.~\ref{fig2}. 
}
\end{figure}

It is interesting to note that a circular structure,
formed by eigenvalues $\lambda_i$ with $|\lambda_i|$ being close to unity
(see red and blue point in top left and right panels
of Fig.~\ref{fig3}), is rather similar to those appearing in the
Ulam networks of intermittency maps studied in \cite{ulamnet2}
(see Fig.4 there). 
Following an analogy with the dynamics of these one-dimensional maps
we may say that the eigenstates related to such a circular structure
corresponds to quasi-isolated communities, being similar to orbits in  
a vicinity of intermittency region,
where the information circulates mainly inside the community
with only a very little flow outside of it. 

The eigenstates of $G$ and $G^*$ with $|\lambda|$ being unity or close to unity
are shown in Fig.~\ref{fig3}. 
For the PageRank $P$ (CheiRank $P^*$) we compare its dependence 
on the corresponding index $K$ ($K^*$) with the PageRank (CheiRank) 
of the Wikipedia network
analyzed in \cite{wiki,univpr,2dmotor} which size $N$ (number of links 
$N_\ell$) is by a factor of 10 (20) smaller.
Surprisingly we find that the PageRank
$P(K)$ of Twitter, approximated by the algebraic decay
$P(K)=a/K^\beta$, has a slower drop as compared to Wikipedia case.
Indeed, we have $\beta = 0.540 \pm  0.004$ ($a=0.00054 \pm 0.00002$)
for the PageRank of 
Twitter in the range 
$1 \leq \log_{10} K \leq 6$ (similar value 
as in \cite{vivek} for the range $ \log_{10} K \leq 5.5$)
while we have  $\beta = 0.767 \pm 0.0005$ ($a=0.0086 \pm 0.00035$)
for the same range of PageRank of
Wikipedia network.
Also we have a sharper drop of CheiRank 
with $\beta=0.857 \pm 0.003$ ($a=0.0148 \pm 0.0004$)
compared to those of PageRank of Twitter
while for  CheiRank of Wikipedia network we
find an opposite tendency 
($\beta= 0.620 \pm 0.001, a= 0.0015 \pm 0.00002$) in the same index range.
Thus for Twitter network the PageRank is more delocalized compared to
CheiRank (e.g. $P(1) < P^*(1)$) while usually one has
the opposite relation (e.g. for Wikipedia $P(1) > P^*(1)$).
We attribute this to the enormously high inter-connectivity between
the top PageRank nodes $K \leq 10^4$ which is well visible 
in Fig.~\ref{fig1}.

We should also point out a specific property of PageRank and CheiRank vectors
which has been already noted in \cite{integers}: there are some 
degenerate plateaus
in $P(K(i))$ or $P^*(K^*(i))$ with absolutely the same 
values of $P$ or $P^*$ for a few nodes. 
For example, for the Twitter network we
have the appearance of the first 
degenerate plateau at $P=7.639 \times 10^{-7}$ for 
$196489 \leq K \leq 196491$. 
As a result the PageRank index $K$ 
can be ordered in various ways. We attribute this phenomenon to 
the fact that the matrix elements of $G$ are composed from
rational elements that leads to such type of degeneracy.
However, the sizes of such degenerate plateaus are
relatively short and they do not influence significantly the PageRank order.
Indeed, on large scales the curves of $P(K)$, $P^*(K^*)$
are rather smooth being characterized by a finite slope
(see Fig,~\ref{fig3}). Similar type of degenerate plateaus
exits for networks of Wikipedia, Cambridge and Oxford Universities.

Other eigenvectors of $G$ and $G^*$ of Twitter network
are shown by color curves in Fig.~\ref{fig3}. We see that the 
shape of eigenstates with $\lambda_1$ and $\lambda_2$, 
shown as a function of their monotonic decrease index $K_i$,
is well pronounced in $P(K)$. Indeed, these vectors
have a rather small gap separating them from unity
($|\Delta \lambda| \sim 2 \times 10^{-5}$) 
and thus they significantly contribute to the PageRank
at $\alpha=0.85$. At the same time we note that the gap
values are significantly smaller than those for certain British Universities
(see e.g. Fig.4 in \cite{univpr}). We argue that a larger
number of links $\xi_\ell$ for Twitter is at the origin
of moderate spectral gap between the core space spectrum and $\lambda=1$.
The eigenvectors of $G^*$ have less slope variations 
and their decay is rather similar to the decay of CheiRank vector $P^*(K^*)$.

\begin{figure}[h]
\begin{center}
\includegraphics[width=0.48\textwidth]{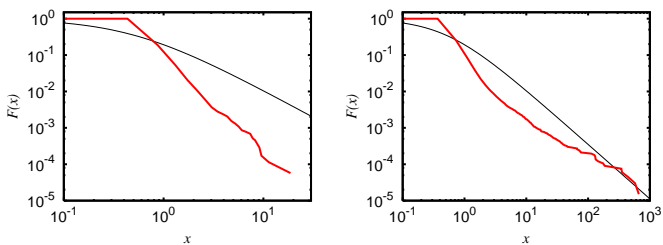}
\end{center}
\caption{\label{fig4} 
Fraction of invariant subspaces $F$ with dimensions larger than $d$ 
as a function of the rescaled variable $x=d/\langle d\rangle$,
where $\langle d\rangle$ is the average subspace dimension. Left 
(right) panel corresponds to the matrix $S$ ($S^*$) for the Twitter network 
(thick red curve) with $\langle d\rangle=2.30$ ($2.72$). 
The tail can be fitted for $x\ge 0.5$ ($x\ge 10$) 
by the power law $F(x)=a/x^{b}$ with 
$a=0.092\pm 0.011$ and $b=2.60\pm 0.07$
($a=0.0125\pm 0.0008$ and $b=0.94\pm 0.02$).
The thin black line is $F(x)=(1+2x)^{-1.5}$ which corresponds to the 
universal behavior of $F(x)$ found in Ref.~\cite{univpr} for the WWW of British 
university networks. 
}
\end{figure}

Finally, in Fig.~\ref{fig4} we use the approach developed in \cite{univpr}
and analyze the dependence of the fraction of invariant subspaces $F(x)$ 
with dimensions larger than $d$ 
on the rescaled variable $x=d/\langle d\rangle$ where $\langle d\rangle$ is 
the average subspace dimension. In \cite{univpr} it was found that
the British University networks are characterized by a universal
functional distribution $F(x)=1/(1+2x)^{3/2}$. For the Twitter network
we find significant deviations from such a dependence
as it is well seen in Fig.~\ref{fig4}. The tail can 
be fitted by the power law $F(x)\sim x^{-b}$ with the exponent
$b=2.60$ for $G$ and $b=0.94$ for $G^*$.
It seems that with the increase of number of links per node $\xi_\ell$
we start to see deviations from the above universal distribution:
it is visible for Wikipedia network (see Fig.7 in \cite{univpr})
and becomes even  more pronounced for the Twitter network.
We assume that a large value of  $\xi_\ell$ for Twitter leads to a
change of the percolation properties of the network
generating other type of distribution $F$
which properties should be studied in more detail in further.

%%%%%%%%%%%%%%%%%%%%%%%%%%%%%%%%%%%%%%%%%%%%%%%%%%%%%%%%%
\section{CheiRank versus PageRank of Twitter}
\label{sec4}

\begin{figure}[h]
\begin{center}
\includegraphics[width=0.48\textwidth]{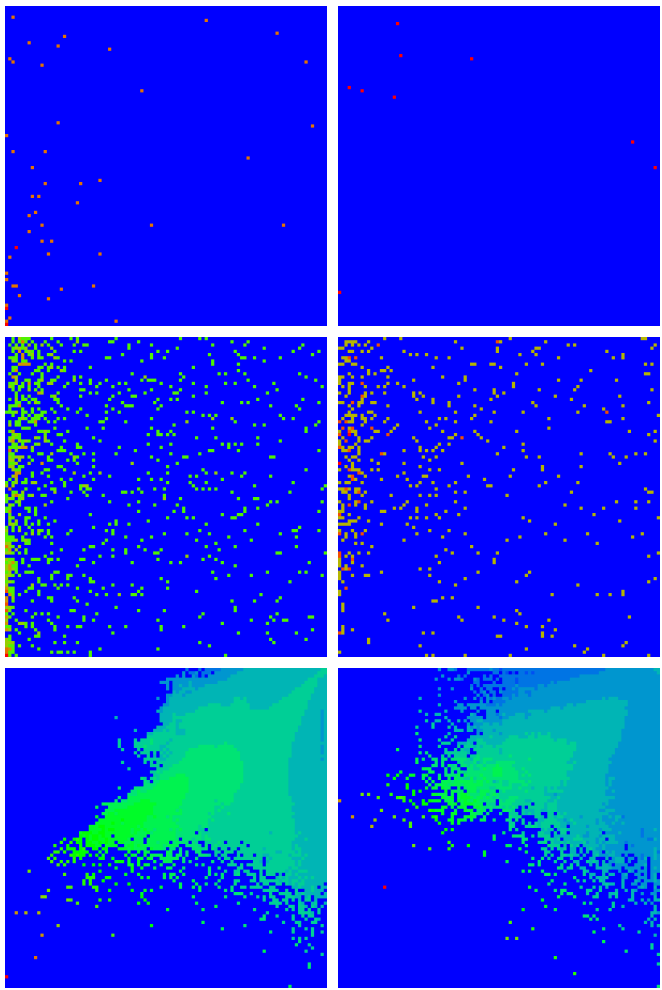}
\end{center}
\caption{\label{fig5} Density of nodes  $W(K,K^*)$ on PageRank-CheiRank plane $(K,K^*)$
for Twitter (left panels) and Wikipedia (right panels).
Top panels show density in the range $1 \leq K,K^* \leq 1000$
with averaging over cells of size $10\times10$;
middle panels show the range  $1 \leq K,K^* \leq 10^4$
with averaging over cells of size $100\times100$;
bottom panels show density averaged over $100\times100$ 
logarithmically equidistant grids for $0 \leq \ln K, \ln K^* \leq \ln N$,
the density is averaged over all nodes inside each cell of the grid,
the normalization condition is $\sum_{K,K^*}W(K,K^*)=1$.
Color varies from blue at zero value to red at maximal density value.
At each panel the $x$-axis corresponds to $K$ (or $\ln K$ for the 
bottom panels) 
and the $y$-axis to $K^*$ (or $\ln K^*$ for the bottom panels). 
}
\end{figure}

As discussed in \cite{alik,wiki,2dmotor}
each network node $i$ has its own PageRank index $K(i)$
and CheiRank index $K^*(i)$ and, hence, the ranking of network nodes 
becomes a two-dimen\-sional (2DRanking).
The distribution of Twitter nodes in the PageRank-CheiRank plane
$(K,K^*)$ is shown in Fig.~\ref{fig5} (left column)
in comparison to the case of the Wiki\-pedia network 
from \cite{wiki,2dmotor} (right column).
There are much more nodes inside the square of size 
$K,K^* \leq 1000$ for Twitter as compared to the case of Wikipedia.
For the squares of  larger sizes the densities become comparable.
The global logarithmic density distribution is shown in 
the bottom panels of Fig.~\ref{fig5} for both networks.
The two densities have certain similarities in their
distributions: both have a maximal density along a certain ridge
along a line $\ln K^* =\ln K +\,$const. However, for the Twitter network
we have a significantly larger number of nodes
at small values $K,K^*<1000$ while in the Wikipedia network
this area is practically empty.

\begin{figure}[h]
\begin{center}
\includegraphics[width=0.48\textwidth]{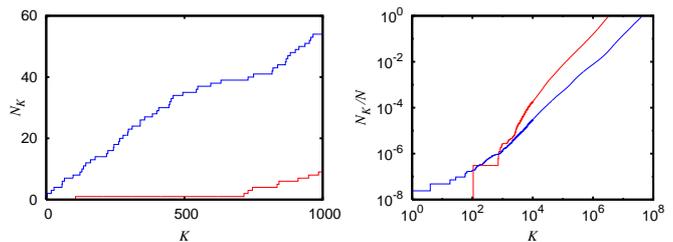}
\end{center}
\caption{\label{fig6} 
Dependence of number of nodes $N_K$, counted
inside the square of size $K \times K$ on PageRank-CheiRank plane,
on $K$ for Twitter (blue curve) and Wikipedia (red curve);
left panel shows data for $1 \leq K \leq 1000$ in linear scale,
right panel shows data in log-log scale for the whole range of $K$.
}
\end{figure}

The striking difference between the Twitter and Wiki\-pedia networks is 
in the number of points $N_K$,
located inside a square area of size $K\times K$ in the 
PageRank-CheiRank plane. This
is directly illustrated in Fig.~\ref{fig6}: at $K=500$ 
there are 40 times more nodes for Twitter,
at $K=1000$ we have this ratio around 6. We note that a similar dependence
$N_K$ was studied in \cite{2dmotor} for Wikipedia, British Universities 
and Linux Kernel networks (see Fig.8 there),
where in all cases the initial growth of $N_K$ was significantly smaller 
as compared to the Twitter network considered here. 

\begin{figure}[h]
\begin{center}
\includegraphics[width=0.48\textwidth]{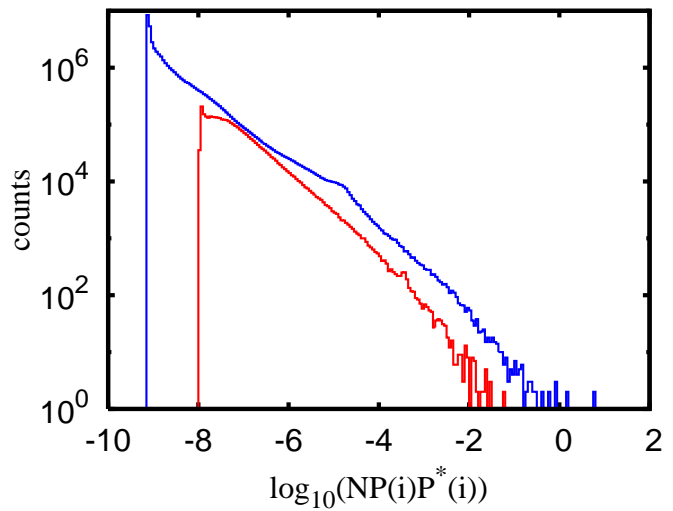}
\end{center}
\caption{\label{fig7} 
Histogram of frequency appearance of   
correlator components $\kappa_i=N P(K(i)) P^*(K^*(i))$ 
for networks of Twitter (blue) and Wikipedia (red).
For the histogram 
the whole interval  $10^{-{10}} \leq \kappa_i \leq 10^2$ is divided in 
240 cells of equal size in logarithmic scale.
}
\end{figure}

Another important characteristics of 2DRanking is the correlator $\kappa$
(\ref{eq2}) between PageRank and CheiRank vectors. We find for Twitter
the value $\kappa=112.60$ which is by a factor 30 - 60 larger
compared to this value for Wikipedia (4.08),
Cambridge and Oxford University networks of 2006 
considered in \cite{wiki,univpr,2dmotor}. The origin of such a large value of
$\kappa$ for the Twitter network becomes more clear from
the analysis of the distribution
of individual node contributions $\kappa_i=N P(K(i)) P^*(K^*(i))$ in 
the correlator sum (\ref{eq2}) shown in Fig.~\ref{fig7}.
We see that there are certain nodes with very large $\kappa_i$ values
and even if there are only few of them still they
give a significant contribution to the total correlator value.
We note that there is a similar feature for the Cambridge University network in 2011
as discussed in \cite{2dmotor} even if there one finds a smaller value 
$\kappa=30$. Thus we see that for certain nodes we have strongly correlated
large values of $P(K(i))$ and $P^*(K^*(i))$ explaining the largest 
correlator value $\kappa$ among all networks studied up to now.
We will argue below that this is related to a very strong inter-connectivity
between top $K$ PageRank  users of the Twitter network.

%%%%%%%%%%%%%%%%%%%%%%%%%%%%%%%%%%%%%%%%%%%%%%%%%%%%%%%%%
\section{Discussion}
\label{sec5}

\begin{figure}[h]
\begin{center}
\includegraphics[width=0.48\textwidth]{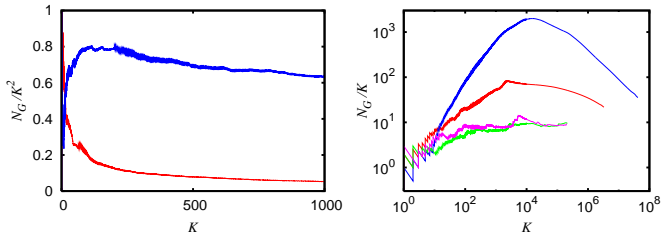}
\end{center}
\caption{\label{fig8} 
Left panel: dependence of the 
area density $g_K=N_G/K^2$  of nonzero elements of the adjacency
matrix among top PageRank nodes on the PageRank index $K$
for Twitter (blue curve) and Wikipedia (red curve) networks,
data are shown in linear scale.
Right panel: linear density $N_G/K$ 
of same matrix elements shown for the whole range of $K$ in log-log scale
for Twitter (blue curve), Wikipedia (red curve), Oxford University 2006 
(magenta curve)
and Cambridge University 2006 (green curve) (curves from top to bottom
at $K=100$).
}
\end{figure}

In this work we study the statistical properties of the Google matrix
of Twitter network including its spectrum, eigenstates and 2DRanking
of PageRank and CheiRank vectors.
The comparison with Wikipedia shows that for Twitter we have much stronger 
correlations between Page\-Rank and CheiRank vectors. 
Thus for the Twitter network there are nodes which are very well known
by the community of users and at the same time they are very communicative
being strongly connected with top Page\-Rank nodes.
We attribute the origin of this phenomenon to
a very strong connectivity between top $K$ nodes for Twitter as 
compared to the Wikipedia network. 
This property is illustrated in Fig.~\ref{fig8}
where we show the number of nonzero elements $N_G$ of the Google matrix,
taken at $\alpha=1$ and counted in the top left corner
with indexes being smaller or equal to $K$ 
(elements in columns of dangling nodes are not taken into account).
We see that for $K \leq 1000$ we have for Twitter the 2D density 
of nonzero elements
to be on a level of 70\% while for Wikipedia this density is 
by a factor 10 smaller.
For these two networks the dependence of $N_G$ on $K$ 
at $K\leq 1000$ is well described by a power law
$N_G=a N^b$ with $a=0.72 \pm 0.01$, $b=1.993 \pm 0.002$ for Twitter
and $a=2.10 \pm 0.01$, $b=1.469 \pm 0.001$ for Wikipedia.
Thus for Twitter the top $K \leq 1000$ elements fill about $70\%$ of the matrix
and about $20\%$ for size $K \leq 10^4$. For Wikipedia the filling factor
is smaller by a factor $10 - 20$. An effective number of links per node
for top $K$ nodes is given by the ratio
$N_G/K$ which is equal to $\xi_\ell$ at $K=N$. The dependence of this ratio 
on $K$ is shown in Fig.~\ref{fig8} in right panel. We see a striking difference
between Twitter network and networks of Wikipedia, 
Cambridge and Oxford Universities.
For Twitter the maximum value of $N_G/K$ is 
by two orders of magnitude larger as compared to
the Universities networks, and by a factor 20 larger than for Wikipedia.
Thus the Twitter network is characterized by a very strong connectivity
between top PageRank nodes which can be considered 
as the Twitter elite \cite{vivek}. 

It is interesting to note that for $K \leq 20$
the Wikipedia network has a larger value of the ratio
$N_G/K^2$ compared to the Twitter network, but the situation
is changed for larger values of $K>20$. In fact the first top 20 
nodes of Wikipedia network are mainly composed from world countries
(see \cite{wiki}) which are strongly interconnected due to
historical reasons. However, at larger values of $K$
Wikipedia starts to have articles on various subjects
and the ratio $N_G/K^2$ drops significantly. On the other hand, 
for the Twitter network we see that a large group of 
very important persons (VIP) 
with $K < 10^4$ is strongly interconnected.
This dominant VIP structure has certain  similarities with the
structure of transnational corporations and their ownership network
dominated by a small tightly-knit core of 
financial institutions \cite{worldcompany}.
The existence of a solid phase of industrially devoloped,
strongly linked countries is also established for the world trade
network obtained from the United Nations COMTRADE data base \cite{wtn}.
It is possible that such super concentration
of links between top Twitter users
results from a global increase of number of links per node
characteristic for such type of social networks.
Indeed, the recent analysis of the Facebook network
shows a significant decrease of degree of separation
during the time evolution of this network \cite{vignaface}.
Also the number of friendship links per node
reaches as high value as $\xi_\ell \approx  100$
at the current Facebook snapshot studied in \cite{vignaface}
(see Table 2 there). This significant growth of $\xi_\ell$
during the time evolution of social networks
leads to an enormous concentration of links 
among society elite at top PageRank users
and may significantly influence the process of strategic decisions
on such networks in the future. 
The growth of $\xi_\ell$ leads also to a significant 
decrease of the exponent $\beta$ of algebraic 
decay of PageRank which is known to be
$\beta \approx 0.9$ for the WWW 
(see e.g. \cite{donato,upfal,meyerbook})
while for the Twitter network we find
$\beta \approx 0.5$ (see also \cite{vivek}). 
This tendency may be a precursor of a delocalization transition
of the PageRank vector emerging at a large values of 
$\xi_\ell$. Such a delocalization would 
lead to a flat PageRank probability
distribution and a strong drop of the 
efficiency of the information retrieval process.
It is known that for the Ulam networks
of dynamical maps such a delocalization indeed takes place 
under certain conditions \cite{ulamnet2,ulamnet1}. 

Our results show that 
the strong inter-connectivity of VIP users
with about top 1000 PageRank indexes 
dominates the information flow on the network.
This result is in line with the recent studies of
opinion formation of the Twitter network \cite{vivek}
showing that the top 1300 PageRank users of Twitter
can impose their opinion for the whole network
of 41 million size.  Thus we think that the 
statistical analysis presented here 
plays a very important role for
a better understanding of decision making 
and opinion formation on the modern social networks.

The present size of the Twitter network
is by a factor 3.5 larger as compared to its size in 2009
analyzed in this work.
Thus it would be very interesting to
extend the present analysis to the
current status of the Twitter network
which now includes all layers of the world society.
Such an analysis will allow to understand in 
an better way the process of information 
flow and decision making on social networks.

This work is supported in part by the EC FET Open project 
``New tools and algorithms for directed network analysis''
(NADINE $No$ 288956). We thank S.Vigna for 
providing us a friendly access
to the Twitter dataset \cite{twitter2009,twittervigna}.
We also acknowledge the France-Armenia collaboration grant 
CNRS/SCS $No$ 24943 (IE-017) on ``Classical and quantum chaos''.

%\newpage
%\vfill
%%%%%%%%%%%%%%%%%%%%%%%%%%%%%%%%%%%%%%%%%%%%%%%%%%%%%%%%%

\end{document}